\documentclass[stef,twoside]{stefano}

\usepackage{amsmath}
\usepackage{amssymb}
\usepackage{graphics}
\usepackage{rotating}
\usepackage{cite}
\usepackage{color}

\def\makeheadbox{{%
\hbox to0pt{\vbox{\baselineskip=10dd\hrule\hbox
to\hsize{\vrule\kern3pt\vbox{\kern3pt \hbox{  {\sc Journal of
Mathematical Physics}, {\bf 48}, 025111-10 (2007) } \hbox{ {\sc
{\color{blue}{dma}}[{\color{black}{imecc}}]{\color{red}{UniCamp}}
} \hspace*{10.4cm} {\color{blue}{$\boldsymbol{\Sigma \delta
\Lambda}$}} }
\kern3pt}\hfil\kern3pt\vrule}\hrule}%
\hss}}}

\def\0{\mbox{\tiny $0$}}
\def\1{\mbox{\tiny $1$}}
\def\2{\mbox{\tiny $2$}}
\def\3{\mbox{\tiny $3$}}
\def\4{\mbox{\tiny $4$}}
\def\5{\mbox{\tiny $5$}}
\def\6{\mbox{\tiny $6$}}
\def\7{\mbox{\tiny $7$}}
\def\8{\mbox{\tiny $8$}}
\def\9{\mbox{\tiny $9$}}
\def\I{\mbox{\tiny $I$}}
\def\II{\mbox{\tiny $II$}}
\def\M{\mbox{\tiny max}}
\def\min{\mbox{\tiny min}}
\def\i{\mbox{\tiny inc}}
\def\r{\mbox{\tiny ref}}
\def\t{\mbox{\tiny tra}}
\begin{document}
%
%%%%%%%%%%%%%%%%%%%%%%%%%%%%%%%% PAPER %%%%%%%%%%%%%%%%%%%%%%%%%%%%%%%%%%%%%

\title{\Large QUATERNIONIC WAVE PACKETS}

\author{
Stefano De Leo\inst{1}
%\thanks{Partially supported by the FAPESP grant 99/09008--5.}
\and Gisele C. Ducati\inst{2}
%\and Tiago M. Madureira\inst{2}
%\thanks{Supported by a CAPES PhD fellowship.}
%\and
%Pietro Rotelli\inst{3}
}

\institute{
Department of Applied Mathematics, University of Campinas\\
PO Box 6065, SP 13083-970, Campinas, Brazil\\
{\em deleo@ime.unicamp.br}
 \and
Department of Mathematics, University of Parana\\
PO Box 19081, PR 81531-970, Curitiba, Brazil\\
{\em ducati@mat.ufpr.br}
% {\em tmadureira@mat.ufpr.br}
%\and
%Department of Physics, University of
%Lecce and INFN Lecce\\
%PO BOX 193, CAP 73100, Lecce, Italy\\
%{\em rotelli@le.infn.it}
}

%%%%%%%%%%%%%%%%%%%%%%%%%%%%%%%%%%%%%%%%%%%%%%%%%%%%%%%%%%%%%%%%%%%%%%%%%%%
%%%%%%%%%%%% DATE ABSTRACT PACS % %%%%%%%%%%%%%%%%%%%%%%%%%%%%%%%%%%%%%%%%%

\date{Submitted: {\em November, 2006}. Accepted: {\em April, 2007}.}
% Warning: Where is the date?

\abstract{We compare the behavior of a wave packet in the presence of a
complex and a pure quaternionic potential step. This analysis, done for a
gaussian convolution function, sheds new light on the possibility to
recognize quaternionic deviations from standard quantum mechanics.}

%%%%%%%%%%%%%%%%%%%%%%%%%%%%%%%%%%%%%%%%%%%%%%%%%%%%%%%%%%%%%%%%%%%%%%%
%%%%%%%%%%%%%%%%%%%%%%%%%%%%%%%%%%%%%%%%%%%%%%%%%%%%%%%%%%%%%%%%%%%%%%%

% Warning: No PACS code given

%02.10.Hh Rings and algebras
%02.10.Ud Linear algebra
%02.10.Yn Matrix theory

%02.30.Hq Ordinary differential equations
%02.30.Jr Partial differential equations
%02.30.Tb Operator theory

%Algebraic methods in quantum mechanics, 03.65.Fd
%Scattering theory (quantum mechanics), 03.65.Nk

%Neutrino oscillations, 14.60.Pq

%15A18 Eigenvalues, singular values, and eigenvectors
%15A21 Canonical forms, reductions, classification

%15A30 Algebraic systems of matrices

%15A90 Applications of matrix theory to physics

%47D15 Linear spaces of operators

%47D25 Operator algebras on Hilbert space

%47E05 Ordinary differential operators
%47F05 Partial differential operators

%\offprints{~Stefano De Leo.}

\titlerunning{\sc quaternionic wave packets}

\maketitle

%%%%%%%%%%%%%%%%%%%%%%%%%%%%%%%%%%%%%%%%%%%%%%%%%%%%%%%%%%%%%%%%%%%%%%%%%%%%%%%
%                               INTRODUCTION
%%%%%%%%%%%%%%%%%%%%%%%%%%%%%%%%%%%%%%%%%%%%%%%%%%%%%%%%%%%%%%%%%%%%%%%%%%%%%%%

\section*{I. INTRODUCTION}

This article represents the third work of the authors on the Schr\"odinger
equation in the presence of a quaternionic potential step. In the first
paper\cite{JMP47a}, we have shown that, for such a potential, it is
possible to calculate an analytic plane wave solution. This represents, to
the best of our knowledge,  the first case in which an analytic solution
has been given for a quaternionic quantum mechanical potential problem
(previous plane wave studies, regarding the
barrier\cite{DAV89,DAV92,DEDUNI02} and well\cite{DEDU05} potentials,
required numerical calculations). The possibility to work with an analytic
plane wave solution allowed, in our second paper\cite{JMP47b}, a detailed
discussion of the quaternionic diffusion through the stationary phase
method. The motivation to write a third paper on this topic is mainly due
to the old wish of the authors to find qualitative differences between
complex and quaternionic formulation of quantum mechanics which could be
useful in identifying the evidence of quaternionic potentials in the case
in which such potentials really exist. In this spirit, the quaternionic
diffusion, at the moment discussed from a general point of view  by using
the stationary phase method, should be investigated by directly analyzing
the motion of the incident, reflected and transmitted wave packets in a
potential step problem. The numerical results of this study  can be then
interpreted by looking for analytical approximations of the incident,
reflected and transmitted wave packets in the case of gaussian convolution
functions. The comparison between the complex and pure quaternionic case
for the potential step seems to be the best starting point to analyze
qualitative differences between quaternionic and complex quantum mechanics.
Indeed, more complicated potentials  can often been seen as successive
potential steps. For example, a potential barrier can be successful studied
as a two-step problem\cite{MPLA19,EPJC46}. The investigation proposed in
this paper could also give a final answer to the old question concerning
the possibility to fit a pure quaternionic potential by a complex one a
consequently to never recognize quaternionic deviations from standard
quantum mechanics.

\section*{II. PLANE WAVE ANALYSIS}

Before to give the explicit plane wave solution for a pure quaternionic
potential step, let us show how, by a simple re-phasing of the wave
function, we can transform, without lost of generalization, the
$(j,k)$-part of the quaternionic potential
\[ i\,V_{\1}+j\,V_{\2}+k\,V_{\3}\,\,,\]
into a pure $j$-part, i.e.
\[ i\,V_{\1}+j\,\sqrt{V_{\2}^{^{\2}}+V_{\3}^{^{\2}}}\,\,.\]
This will greatly simplify our presentation. Let us begin by considering
the quaternionic Schr\"odinger equation\cite{ADL}
\begin{equation}
\label{sceq1} i\,\frac{\hbar^{^{\2}}}{2m}\,\Phi_{xx}(x,t)-\left(\,i\,V_{\1}
+j\,V_{\2}+ k\, V_{\3}\,\right)\,\Phi(x,t) =  \hbar\,\Phi_t(x,t)\, \,.
\end{equation}
This partial differential equation can be reduced, by using the well-know
separation of variables
\[
\Phi(x,t) = \varphi(x)\, \exp[-\,i\,E\,t/\hbar]\,\, ,
\]
to the following second order ordinary differential equation\cite{DEDUC01}
\begin{equation}
\label{sceq2} i\,\frac{\hbar^{^{\2}}}{2m}\,\varphi''(x)-\left(\,i\,V_{\1}
+j\,V_{\2}+ k\, V_{\3}\, \right)\,\varphi(x) = - \, \varphi(x)\,E\, i\,\,.
\end{equation}
By multiplying (from the left) the previous equation by an unitary complex
number  $e^{i\alpha}$ and re-writing the pure quaternionic part of the
potential as
\[ j\,V_{\2}+ k\, V_{\3}=j\,
\sqrt{V_{\2}^{^{\2}}+V_{\3}^{^{\2}}} \,\,e^{-i\theta}\,\, ,\] where
$\theta=\arctan \left[V_{\3}/V_{\2}\right]$, we find
\begin{equation}
e^{i\alpha}\,\left[\,
i\,\frac{\hbar^{^{\2}}}{2m}\,\varphi''(x)-\left(\,i\,V_{\1}
+j\,\sqrt{V_{\2}^{^{\2}}+V_{\3}^{^{\2}}} e^{-i\theta}
\right)\,\varphi(x)\,\right] = - \,e^{i\alpha}\, \varphi(x)\,E\, i\,\,.
\end{equation}
By observing that $e^{i\alpha}j=j\,e^{-i\alpha}$, we can rewrite the
previous equation as follows
\begin{equation}
i\,\frac{\hbar^{^{\2}}}{2m}\,\left[e^{i\alpha}\,\varphi(x)\right]''-
i\,V_{\1}\, \left[e^{i\alpha}\,\varphi(x)\right]-
j\,\sqrt{V_{\2}^{^{\2}}+V_{\3}^{^{\2}}}\,\left[e^{-i(\alpha+\theta)}\,
\varphi(x)\right]  = - \,\left[e^{i\alpha}\, \varphi(x)\right]\,E\, i\,\,.
\end{equation}
Consequently, the choice $\alpha=-\,\theta/2$ conduces to
\begin{equation}
\label{sceq3}
 i\,\frac{\hbar^{^{\2}}}{2m}\,\psi''(x)-\left(\,i\,V_{\1} +j\,
 \sqrt{V_{\2}^{^{\2}}+
V_{\3}^{^{\2}}}\, \right)\,\psi(x) = - \, \psi(x)\,E\, i\,\,,
\end{equation}
where
\[ \psi(x)=e^{-i\frac{\theta}{2}}\,\varphi(x)\,\,.\]
It is important to observe that constant phases paly no role in the
stationary phase method and, thus, such a re-phasing has no physical effect
on the motion of the incident, reflected and transmitted wave packets. For
the diffusion problem, $E > V_{\0}$, we shall compare the complex case,
$V_{\0}=V_{\1}$ ($V_{\2}=V_{\3}=0$), with the pure quaternionic case,
$V_{\0}=\sqrt{V_{\2}^{^{\2}}+V_{\3}^{^{\2}}}$ ($V_{\1}=0$).

\subsection*{II.A COMPLEX POTENTIAL STEP}
By setting $V_{\1}=V_{\0}$ and $V_{\2}=V_{\3}=0$ in Eq.(\ref{sceq3}), we
find the standard Schr\"odinger equation,
\begin{equation}
i\,\frac{\hbar^{^{\2}}}{2m}\,\psi_c''(x)- i\,V_{\0} \,\psi_c(x) =
  - \, \psi_c(x)\,E\, i\,\,,
\end{equation}
whose analytic plane wave solution is given by \cite{COHEN}
\begin{equation}
\label{solc}
\begin{array}{llcl}
\mbox{\sc [I] free region ($x<0$) :} \hspace*{.5cm}  &
\psi_{c,\I}(\epsilon,x) & = &
\exp[\,i\,\epsilon\,x\,] + r_c(\epsilon) \, \exp[\,-\,i\, \epsilon \,x\,] \,\, , \\ \\
\mbox{\sc [II] potential region ($x>0$) :} \hspace*{.5cm}  &
\psi_{c,\II}(\epsilon,x) & = & t_c(\epsilon)\,\exp[\,i\,\sigma\, x\,] \,\,
,
\end{array}
\end{equation}
where
\[
\epsilon   =  \sqrt{2m\,E}/ \hbar \,\,\,,\,\,\,\,\,\sigma
=\sqrt{2m\,(E-V_{\0})}/ \hbar \,\, ,
\]
and
\[
r_c(\epsilon)=(\epsilon - \sigma)/(\epsilon + \sigma)\,\,\, , \, \,\,\,\,
t_c(\epsilon) = 2\, \epsilon /(\epsilon + \sigma) \,\, .
\]

\subsection*{II.B PURE QUATERNIONIC POTENTIAL STEP}
By setting in Eq.(\ref{sceq3}) $V_{\1}=0$ and
$\sqrt{V_{\2}^{^{\2}}+V_{\3}^{^{\2}}}=V_{\0}$, we obtain
\begin{equation}
 i\,\frac{\hbar^{^{\2}}}{2m}\,\psi_q''(x)- j\,
 V_{\0}\,\psi_q(x) = - \, \psi_q(x)\,E\, i\,\,.
\end{equation}
The analytic plane wave solution for the pure quaternionic potential step
reads
\begin{equation}
\label{solq}
\begin{array}{llcl}
\mbox{\sc [I] free region :} \hspace*{.05cm}  & \psi_{q,\I}(\epsilon,x) & =
& \exp[\,i\,\epsilon\,x\,] + r_q(\epsilon) \, \exp[\,-\,i\, \epsilon \,x\,]
+
j\, \tilde{r}_q(\epsilon)\, \exp[\,\epsilon\, x\,]\,\, , \\ \\
\mbox{\sc [II] potential region  :} \hspace*{.05cm}  &
\psi_{q,\II}(\epsilon,x) & = & (1+j\, w)\, t_q(\epsilon)\,\exp[\,i\,\rho\,
x\,] + (\bar{w}+j)\, \tilde{t}_q(\epsilon)\,\exp[\,-\,\rho\,x\,]\,\, ,
\end{array}
\end{equation}
where
\[
\epsilon   =  \sqrt{ 2m\,E}/\hbar \,\,\,,\,\,\,\,\,\rho= \sqrt{ 2m\,
\sqrt{E^{^{\2}}-V_{\0}^{^{\2}}}}/\hbar \,\,\, ,\,\,\,\,\, w = -\, i
\,V_{\0}\,\mbox{\Large $/$}\left( E +
\sqrt{E^{^{\2}}-V_{\0}^{^{\2}}}\right)\,\,,
\]
and
\[
\begin{array}{lcl}
r_q(\epsilon) & = & (\epsilon - \rho) \,\exp\left[\,i\,\arctan
\left(\epsilon/\rho\right)\,\right]/\sqrt{\epsilon^{\2}+\rho^{\2}} \,\, ,\\
\tilde{r}_q(\epsilon) & = & (1+i)\,\epsilon\,w/(\epsilon+\rho)\, \,\,,\\
t_q(\epsilon) & = & \epsilon/\rho \,\, ,\\
\tilde{t}_q(\epsilon) & = &
\sqrt{\epsilon^{\2}+\rho^{\2}}\,\epsilon\,w\,\exp\left[\,-\,i\,\arctan
\left(\rho/\epsilon\right)\,\right]/\left[\rho\left(\epsilon+\rho\right)\right]
 \,\,\,.
\end{array}
\]
For a detailed derivation of the plane wave solution for a quaternionic
potential step, we refer the reader to the paper cited in
ref.\cite{JMP47a}.

\section*{III. WAVE PACKET ANALYSIS}

Until now, we have been concerned only with plane waves. In this section,
we are going to study the time evolution of quaternionic wave packets and
deducing from them several important results. The principle of
superposition guarantees that every {\em real} linear combination of the
plane waves in region I and region II,
\[ \Psi_{\I,\II}(x,t)= \int \mbox{d}\epsilon
\,\,g(\epsilon)\,\,\psi_{\I,\II}(\epsilon,x)
\exp[-\,i\,E\,t/\,\hbar]\,\,\,\,\,\,\,\,
\,\,[E=(\hbar\,\epsilon)^{\2}/\,2\,m]\,\,,\]
 will satisfy the Schr\"odinger equation in the
presence of a potential step. The reason why the use of the wave packet
formalism is very interesting lies in the fact that the calculations, in
the case of a gaussian convolution function
\[g(\epsilon)=\exp\left[a^{\2}(\epsilon-\epsilon_{\0}^{\2})/4\right]/\,2
\sqrt{\pi}\,\,,\]
 can be analytically approximated. This will allow to check and interpret our
numerical results. In the next subsections, we first discuss the wave
packets motion in the case of a standard (complex) potential step and,
then, we analyze the special case of a pure quaternionic potential.

\subsection*{III.A COMPLEX CASE}
The wave packets in region I and region II are given by
\begin{eqnarray*}
 \Omega_{c,\I}(x,t) & = &
\int_{a\,\epsilon_{\min}}^{+\,\infty} \hspace*{-.2cm}\mbox{d}(a\,\epsilon)
\,\,g(\epsilon)\, \left\{\,\exp[\,i\,\epsilon\,x\,] + r_c(\epsilon) \,
\exp[\,-\,i\,\epsilon \,x\,] \,\right\} \,\exp[\,-\,i\, \epsilon^{\2} \,
\hbar \,t/\,2m\,]\,\,,\\
\Omega_{c,\II}(x,t)  & =  & \int_{a\,\epsilon_{\min}}^{+\,\infty}
\hspace*{-.2cm}\mbox{d}(a\,\epsilon)
 \,\,g(\epsilon)\,\left\{\,
t_c(\epsilon)\,\exp[\,i\,\sigma\, x\,] \,\right\}\, \exp[\,-\,i\,
\epsilon^{\2} \, \hbar \,t/\,2m\,] \,\,,
\end{eqnarray*}
where
\[a\,\epsilon_{\min}=a\sqrt{2m\, V_{\0}}/\,\hbar\,\,. \]
We find three wave packets: incident, reflected and transmitted,
\begin{eqnarray*}
\Omega_{c,\i}(x,t) & = & \int_{a\,\epsilon_{\min}}^{+ \,\infty}
\hspace*{-.2cm}\mbox{d}(a\,\epsilon) \,\,g(\epsilon)\,
\exp[\,i\,(\epsilon\,x - \epsilon^{\2} \,
\hbar \,t/\,2m)\,]\,\,,\\
\Omega_{c,\r}(x,t) &=& \int_{a\,\epsilon_{\min}}^{+ \,\infty}
\hspace*{-.2cm}\mbox{d}(a\,\epsilon) \,\,g(\epsilon)\, \, r_c(\epsilon) \,
\exp[\,-\,i\,(\epsilon\,x + \epsilon^{\2} \,
\hbar \,t/\,2m)\,] \,\,,\\
\Omega_{c,\t}(x,t)  & =& \int_{a\,\epsilon_{\min}}^{+ \,\infty}
\hspace*{-.2cm}\mbox{d}(a\,\epsilon)
 \,\,g(\epsilon)\,\,
t_c(\epsilon)\,\exp[\,i\,(\sigma\, x - \epsilon^{\2} \, \hbar \,t/\,2m)\,]
\,\,.
\end{eqnarray*}
The choice of a gaussian convolution function $g(\epsilon)$ peaked in
$\epsilon_{\0}$ ($\epsilon_{\0}>\epsilon_{\min}$) and whose value is
practically zero near to $\epsilon_{\min}$, i.e. $g(\epsilon_{\min})\approx
0$, allows to legitimately approximate the incident wave packet as follows
\begin{eqnarray}
\label{cinc} \Omega_{c,\i}(x,t) & \approx & \int_{- \,\infty}^{+\,\infty}
\hspace*{-.2cm}\mbox{d}(a\,\epsilon) \,\,g(\epsilon)\,
\exp[\,i\,(\epsilon\,x - \epsilon^{\2} \, \hbar \,t/\,2m)\,]
\nonumber \\
 & = & \exp[\,i\,(\epsilon_{\0}x-a^{\2}\epsilon_{\0}^{\2} \tau /2)\,-\,i\,
\arctan(2\tau)/2\,]\, \frac{\exp\left[ -
\left(x/a-a\epsilon_{\0}\tau\right)^{\2}/\,(1+2\,i\,\tau)
\right]}{\left(1+4\,\tau^{\2}\right)^{\1/\4}}\,\,,
\end{eqnarray}
where $\tau=\hbar\,t/m\,a^{\2}$. In the case in which the variation of
$r_c(\epsilon)$ can be neglected compared to that of $g(\epsilon)$, the
reflected wave  has the same form as the incident wave packet\cite{COHEN},
\begin{eqnarray}
\label{cref} \Omega_{c,\r}(x,t) & \approx & r_c(\epsilon_{\0})\,
\int_{a\,\epsilon_{\min}}^{+\,\infty} \hspace*{-.2cm}\mbox{d}(a\,\epsilon)
\,\,g(\epsilon)\, \exp[\,-\, i\,(\epsilon\,x + \epsilon^{\2} \, \hbar
\,t/\,2m)\,]
\nonumber \\
 & = & r_c(\epsilon_{\0})\,\Omega_{c,\i}(-\,x,t)\,\,.
\end{eqnarray}
For the transmitted wave packet, by neglecting the variation of
$t_c(\epsilon)$ compared to that of $g(\epsilon)$ and by approximating
$\sigma$ by its first order Taylor expansion, $\sigma_{\0} + (\epsilon -
\epsilon_{\0})\, \epsilon_{\0}/ \sigma_{\0}$, we find
\begin{eqnarray}
\label{ctra} \Omega_{c,\t}(x,t) & \approx & t_c(\epsilon_{\0})\,
\exp[\,i\,(\sigma_{\0}^{\2}-\epsilon_{\0}^{\2})\,x/\sigma_{\0}\,]
\,\int_{a\,\epsilon_{\min}}^{+\,\infty}
\hspace*{-.2cm}\mbox{d}(a\,\epsilon) \,\,g(\epsilon)\,
\exp[\,i\,(\epsilon\,\epsilon_{\0}\, x/\sigma_{\0} - \epsilon^{\2} \, \hbar
\,t/\,2m)\,]
\nonumber \\
 & = & t_c(\epsilon_{\0})\,
\exp[\,i\,(\sigma_{\0}^{\2}-\epsilon_{\0}^{\2})\,x/\sigma_{\0}\,]
\,\,\Omega_{c,\i}(\epsilon_{\0}\, x/\sigma_{\0},t)\,\,.
\end{eqnarray}
The reflection probability (the ratio between the probabilities of finding
the particle in the reflected packet, at a positive time $t_{\0}$, and in
the incident packet, at time $-\,t_{\0}$) is given by
\begin{eqnarray}
P_{c,\r}& =&\int_{- \,\infty}^{\,0}
\hspace*{-.2cm}\mbox{d}x\,|\,\Omega_{c,\r}(x,t_{\0})\,|^{^{\2}}\,\,\mbox{\Large
$/$}\, \int_{- \,\infty}^{\,0}
\hspace*{-.2cm}\mbox{d}x\,|\,\Omega_{c,\i}(x,-\,t_{\0})\,|^{^{\2}} \nonumber \\
 & \approx & \,|r_c(\epsilon_{\0})|^{^{\2}}\, \int_{- \,\infty}^{\,0}
\hspace*{-.2cm}\mbox{d}x\,|\,\Omega_{c,\i}(-\,x,t_{\0})\,|^{^{\2}}\,\,\mbox{\Large
$/$}\, \int_{- \,\infty}^{\,0}
\hspace*{-.2cm}\mbox{d}x\,|\,\Omega_{c,\i}(x,-\,t_{\0})\,|^{^{\2}} \nonumber \\
& = & \,|r_c(\epsilon_{\0})|^{^{\2}}\, \,.
\end{eqnarray}
Similarly, the transmission probability is given by
\begin{eqnarray}
P_{c,\t}& =&\int^{+\,\infty}_{\,0}
\hspace*{-.2cm}\mbox{d}x\,|\,\Omega_{c,\t}(x,t_{\0})\,|^{^{\2}}\,\,\mbox{\Large
$/$}\, \int_{- \,\infty}^{\,0}
\hspace*{-.2cm}\mbox{d}x\,|\,\Omega_{c,\i}(x,-\,t_{\0})\,|^{^{\2}} \nonumber \\
 & \approx & \,|t_c(\epsilon_{\0})|^{^{\2}}\, \int^{+ \,\infty}_{\,0}
\hspace*{-.2cm}\mbox{d}x\,|\,\Omega_{c,\i}(\epsilon_{\0}\,
x/\sigma_{\0},t_{\0})\,|^{^{\2}}\,\,\mbox{\Large $/$}\, \int_{-
\,\infty}^{\,0}
\hspace*{-.2cm}\mbox{d}x\,|\,\Omega_{c,\i}(x,-\,t_{\0})\,|^{^{\2}} \nonumber \\
& = & \,|t_c(\epsilon_{\0})|^{^{\2}}\,\frac{\sigma_{\0}}{\epsilon_{\0}}\,
\int^{+ \,\infty}_{\,0}
\hspace*{-.2cm}\mbox{d}x\,|\,\Omega_{c,\i}(x,t_{\0})\,|^{^{\2}}\,\,\mbox{\Large
$/$}\, \int_{- \,\infty}^{\,0}
\hspace*{-.2cm}\mbox{d}x\,|\,\Omega_{c,\i}(x,-\,t_{\0})\,|^{^{\2}} \nonumber \\
& = & \,\frac{\sigma_{\0}}{\epsilon_{\0}}\,\,|t_c(\epsilon_{\0})|^{^{\2}}\,
\,.
\end{eqnarray}
It is easy to verify that
\[ P_{c,\r}+P_{c,\t}=\left[\,
\left(\,\frac{\epsilon-\sigma}{\epsilon+\sigma}\,\right)^{^{\2}} +
\frac{\,\sigma}{\,\epsilon}\,\frac{4\,\epsilon^{\2}}{(\epsilon+\sigma)^{^{\2}}}
\,\right]_{\0}=\,1\,\,. \]

\subsection*{III.B PURE QUATERNIONIC CASE}
The wave packets in region I and region II are given by
\begin{eqnarray*}
 \Omega_{q,\I}(x,t) & = &
\int_{a\,\epsilon_{\min}}^{+\,\infty} \hspace*{-.2cm}\mbox{d}(a\,\epsilon)
\,\,g(\epsilon)\, \left\{\,\exp[\,i\,\epsilon\,x\,] + r_q(\epsilon) \,
\exp[\,-\,i\,\epsilon \,x\,] + j\, \tilde{r}_q(\epsilon)\,
\exp[\,\epsilon\,
x\,]\,\right\} \,\exp[\,-\,i\, \epsilon^{\2} \, \hbar \,t/\,2m\,]\,\,,\\
\Omega_{q,\II}(x,t)  & =  &  \int_{a\,\epsilon_{\min}}^{+\,\infty}
\hspace*{-.2cm}\mbox{d}(a\,\epsilon) \,\,g(\epsilon)\,\left\{\,
t_q(\epsilon)\,\exp[\,i\,\rho\, x\,] + \bar{w}\,
\tilde{t}_q(\epsilon)\,\exp[\,-\,\rho\,x\,] \,\right\}\, \exp[\,-\,i\,
\epsilon^{\2} \, \hbar \,t/\,2m\,]  + \nonumber \\
 & & j\, \int_{a\,\epsilon_{\min}}^{\,\infty}
 \hspace*{-.2cm}\mbox{d}(a\,\epsilon)
\,\,g(\epsilon)\,\left\{\, w\, t_q(\epsilon)\,\exp[\,i\,\rho\, x\,] +
\tilde{t}_q(\epsilon)\,\exp[\,-\,\rho\,x\,]\,\right\}\, \exp[\,-\,i\,
\epsilon^{\2} \, \hbar \,t/\,2m\,]\,\,.
\end{eqnarray*}
For sufficiently large negative $t$, only the (complex) incident wave
packet
\[
\Omega_{q,\i}(x,t\ll 0)  =  \int_{a\,\epsilon_{\min}}^{+ \,\infty}
\hspace*{-.2cm}\mbox{d}(a\,\epsilon) \,\,g(\epsilon)\,
\exp[\,i\,(\epsilon\,x - \epsilon^{\2} \, \hbar \,t/\,2m)\,]= \,
\Omega_{c,\i}(x,t) \] exists. For sufficiently large positive $t$, the
evanescent wave packets disappear and we only find one (complex) reflected
wave packet,
\[
\Omega_{q,\r}(x,t\gg 0) = \int_{a\,\epsilon_{\min}}^{+ \,\infty}
\hspace*{-.2cm}\mbox{d}(a\,\epsilon) \,\,g(\epsilon)\, \, r_q(\epsilon) \,
\exp[\,-\,i\,(\epsilon\,x + \epsilon^{\2} \, \hbar \,t/\,2m)\,] \approx
\,r_q(\epsilon_{\0})\, \Omega_{c,\i}(-\,x,t) \,\,,
\]
and one (quaternionic) transmitted wave packet,
\begin{eqnarray*}
\Omega_{q,\t}(x,t\gg 0) &=& \int_{a\,\epsilon_{\min}}^{+ \,\infty}
\hspace*{-.2cm}\mbox{d}(a\,\epsilon) \,\,g(\epsilon)\, \, (1+j\,w)\,
t_q(\epsilon) \, \exp[\,i\,(\rho\,x -  \epsilon^{\2} \, \hbar \,t/\,2m)\,]
\\
&\approx &(1+j\,w_{\0})\,t_q(\epsilon_{\0})\,
\exp[\,i\,(\rho_{\0}^{\4}-\epsilon_{\0}^{\4})\,x/\rho_{\0}^{\3}\,]
\,\,\Omega_{c,\i}(\epsilon_{\0}^{\3}\, x/\rho_{\0}^{\3},t)\,\,.
\end{eqnarray*}
A simple calculation shows that
\begin{equation}
P_{q,\r}  =\int_{- \,\infty}^{\,0}
\hspace*{-.2cm}\mbox{d}x\,|\,\Omega_{q,\r}(x,t_{\0})\,|^{^{\2}}\,\,\mbox{\Large
$/$}\, \int_{- \,\infty}^{\,0}
\hspace*{-.2cm}\mbox{d}x\,|\,\Omega_{q,\i}(x,-\,t_{\0})\,|^{^{\2}} \approx
\,|r_q(\epsilon_{\0})|^{^{\2}}
\end{equation}
and
\begin{equation}
P_{q,\t} = \int^{+\,\infty}_{\,0}
\hspace*{-.2cm}\mbox{d}x\,|\,\Omega_{q,\t}(x,t_{\0})\,|^{^{\2}}\,\,\mbox{\Large
$/$}\, \int_{- \,\infty}^{\,0}
\hspace*{-.2cm}\mbox{d}x\,|\,\Omega_{q,\i}(x,-\,t_{\0})\,|^{^{\2}} \approx
\,\frac{\rho_{\0}^{\3}}{\epsilon_{\0}^{\3}}\,\left(1+|w_{\0}|^{^{\2}}\right)\,
|t_q(\epsilon_{\0})|^{^{\2}}\, \,.
\end{equation}
Finally,
\[ P_{q,\r}+P_{q,\t}=\left[\,
\frac{(\epsilon-\rho)^{^{\2}}}{\epsilon^{\2}+\rho^{\2}} +
\frac{\rho^{\3}}{\epsilon^{\3}}\,\left(\,1+
\frac{\epsilon^{\2}-\rho^{\2}}{\epsilon^{\2}+\rho^{\2}}\,\right)\,
\frac{\epsilon^{\2}}{\rho^{\2}} \right]_{\0}=1\,\,.
\]

\section*{IV. NUMERICAL RESULTS AND INTERPRETATION}

The results of our numerical study are plotted in Fig.\,1, where the
probability densities $|\Omega_c(x,t)|^{^{\2}}$ (complex case) and
$|\Omega_q(x,t)|^{^{\2}}$ (pure quaternionic case) are drawn as a function
of $x/a$ for different values of $\tau=\hbar\,t/ma^{\2}$. The incident wave
packets are centered in $E_{\0}=2\,V_{\0}$ (diffusion) and the potential is
given by $a\,\sqrt{2m\,V_{\0}}=10^{\2}\,\hbar$. The data show an
interesting phenomenon. For the same potential and the same incident
energy, the reflected waves have different amplitudes in the case of a
complex or a pure quaternionic potential step, and the transmitted waves
move with different velocities. The quaternionic transmitted wave packet
moves faster than the complex one. Let us try to understand these results
by using the analytic approximations given in section III for the incident,
reflected and transmitted wave packets and by using the fact that for small
times, $\tau\ll 1$, no spreading effect is present. This allows to simplify
the expressions given in section III for the complex and quaternionic wave
packets. The amplitude of the incident wave is then given by
\[
|\Omega_{q,\i}(x,t)|\approx|\Omega_{c,\i}(x,t)|\approx \exp\left[ -
\left(x/a-a\,\epsilon_{\0}\tau\right)^{\2}\right]
\]
For the reflected waves, we have
\begin{eqnarray*}
|\Omega_{c,\r}(x,t)| & \approx & |r_c(\epsilon_{\0})|\, \exp\left[ -
\left(x/a+\,a\,\epsilon_{\0}\tau\right)^{\2}\right]\,\,, \\
|\Omega_{q,\r}(x,t)| & \approx & |r_q(\epsilon_{\0})|\, \exp\left[ -
\left(x/a+\,a\,\epsilon_{\0}\tau\right)^{\2}\right]\,\,.
\end{eqnarray*}
This implies
\[
\begin{array}{lclcl}
|\Omega_{c,\r}^{^{\M}}(x,t)|^{^{\2}}/\,\,|\Omega_{c,\i}^{^{\M}}(x,t)|^{^{\2}}&
\approx & |r_c(\epsilon_{\0})|^{^{\2}} &
\approx & 2.94\,\times\,10^{-\2}\,\,,\\
|\Omega_{q,\r}^{^{\M}}(x,t)|^{^{\2}}/\,\,|\Omega_{q,\i}^{^{\M}}(x,t)|^{^{\2}}&
\approx & |r_q(\epsilon_{\0})|^{^{\2}} &
\approx & 2.58\,\times\,10^{-\3}\,\,,\\
 |\Omega_{c,\r}(x,t)|^{^{\2}}/\,\,|\Omega_{q,\r}(x,t)|^{^{\2}}&
\approx & |r_c(\epsilon_{\0})|^{^{\2}}/\,\,|r_q(\epsilon_{\0})|^{^{\2}} &
\approx & 11.4\,\,,
\end{array}
\]
which confirms the numerical results given in Fig.1 (negative $x$-axis). We
can also see that the reflected waves move with the same velocity,
\begin{equation}
v_{c,\r}=v_{q,\r}= -\,\hbar\, \epsilon_{\0}/\,m\,\,.
\end{equation}
For the transmitted waves, by observing that
\begin{eqnarray*}
|\Omega_{c,\t}(x,t)| & \approx & |t_c(\epsilon_{\0})|\, \exp\left[ -\,
(\epsilon_{\0}/\sigma_{\0})^{\2}\,\left( x/a -
\,a\,\sigma_{\0}\tau\right)^{\2}
\right]\,\, ,\\
|\Omega_{q,\t}(x,t)| & \approx & |(1+j\,w_{\0})\, t_q(\epsilon_{\0})| \,
\exp\left[ -\, (\epsilon_{\0}/\rho_{\0})^{\6}\,\left( x/a -
\,a\,\rho_{\0}^{\3}\tau/\epsilon_{\0}^{\2}\right)^{\2} \right]\,\,,
\end{eqnarray*}
we find that the centers of the complex and quaternionic transmitted wave
packets move, respectively, with velocities
\begin{equation}
v_{c,\t} = \hbar, \sigma_{\0}/\,m\,\,\,\,\,\,\,\,\mbox{and}\,\,\,\,\,\,\,
v_{q,\t}
 = \hbar\, \rho^{\3}_{\0}/\,m\,\epsilon_{\0}^{\2}\,\, .
\end{equation}
Consequently,
\begin{eqnarray*}
\frac{x^{\M}_{c,\t}(t)}{a} & = & \sqrt{\frac{E_{\0}}{V_{\0}}-1}
\,\,\,\frac{a\sqrt{2m\,V_{\0}}}{\hbar}\,\,
\frac{\hbar \,t}{ma^{\2}} \,\, ,\\
\frac{x^{\M}_{q,\t}(t)}{x^{\M}_{c,\t}(t)} & = &
\left[\,\left(\,\frac{E_{\0}}{V_{\0}}-1\right)\left(\,
\frac{E_{\0}}{V_{\0}}+1\right)^{\3}
\left(\,\frac{V_{\0}}{E_{\0}}\,\right)^{\4} \,\right]^{\1/\4}\,\,,
\end{eqnarray*}
and by using
\[ E_{\0}=2\,V_{\0}\,\,\, ,\,\,\,\,\,\frac{a\sqrt{2m\,V_{\0}}}{\hbar}=10^{^{\2}}\,\,\,, \,\,\, \,\,
\tau= \frac{\hbar \,t}{ma^{\2}} \,\,,\] we find
\begin{eqnarray*}
\frac{x^{\M}_{c,\t}(\tau)}{a} & = & 10^{^{\2}} \tau \,\, ,\\
 \frac{x^{\M}_{q,\t}(\tau)}{x^{\M}_{c,\t}(\tau)} & \approx & 1.14\,\, .
\end{eqnarray*}
This agrees with our numerical calculations as shown by the motion of the
complex and quaternionic transmitted wave packets plotted in Fig.1
(positive $x$-axis). For example, at time $\tau=0.15$ the maximum of the
complex transmitted wave is found at $x=15\,a$, whereas the center of the
quaternionic wave packet (which moves faster) reaches at the same time the
point $x\approx 17.1\, a$.

\section*{V. CONCLUSIONS}

In the last years, the Schr\"odinger equation in the presence of
quaternionic potentials with constant and space-dependent phase has been a
matter of study and discussion in the literature
\cite{DEDUNI02,DEDU05,JMP47b}. Some properties of this class of potentials
are detailed discussed in the Adler book \cite{ADL} which represents a
milestone in such a field. If quaternionic quantum mechanics represents a
possible way to describe the nature, then it becomes relevant to examine
how the predictions of standard theories may be affected by changing from
complex to quaternionic potentials. The first theoretical analysis of
quaternionic potential barriers was developped in \cite{DAV89,DAV92} and
showed that in contrast with the standard complex case where the left and
right transmission probability are equal in magnitude and phase, in the
quaternionic quantum mechanics only the magnitude are equal. So, the
measurement of a phase shift should be an indicator of quaternionic effects
{\em and} space-dependent phase. Nevertheless, as remarked in \cite{ADL},
experiments to detect a phase shift are equivalent to detecting
time-reversal violation and consequently cannot be seen in neutron-optical
experiments \cite{PER79,KAI84,KLE88}. A more complete phenomenology of the
quaternionic potential barrier is given in \cite{DEDUNI02}. With respect to
previous works regarding potential barrier diffusion, in this paper, we
have introduced the quaternionic wave packet formalism. This allow to study
new {\em qualitative} differences between standard quantum mechanics and
theoretical solutions obtained by solving the Schr\"odinger equation in the
presence of a quaternionic step. For a detailed discussion of experimental
proposals on quaternionic potentials and quantum mechanical systems in
which a CP violation is interpreted by potentials with a space-dependent
phase, we refer the reader to references \cite{JMP47a,DEDUNI02}.

In this paper, we have seen that for incoming particles with an energy
spectrum centered in $E_{\0}=2\,V_{\0}$ the complex and quaternionic
transmitted wave packets move with different velocities,
\[\frac{v_{q,\t}}{v_{c,\t}}=\frac{\sqrt{3\sqrt{3}}}{2}\,\,.\] Due to the fact
that the complex and quaternionic incident wave
packets coincide for large negative times (incoming particles), we could
try to fit the motion of the quaternionic wave packet by using a different
complex potential step. For example by choosing a {\em new} complex
potential step, $W_{\0}$, such that
\[ \sqrt{\frac{E_{\0}}{W_{\0}}-1} =\left[\,\left(\,
\frac{E_{\0}}{V_{\0}}\right)^{^{\2}}-1\right]^{\3/\4}
\frac{V_{\0}}{E_{\0}}\,\,,
\]
we find $x^{\M}_{q,\t}(t)=x^{\M}_{c,\t}(t)$. In the particular case
investigated in this paper, $E_{\0}=2\,V_{\0}$, this implies
\[ \frac{E_{\0}}{W_{\0}}=1+\frac{3\sqrt{3}}{4}\,\,.\]
In this scenario, we have the same incident wave and complex or
quaternionic transmitted wave packets which move with the same velocity
(note that the reflected waves, for a same incoming energy spectrum, always
move with the same velocity because propagate in the free potential
region). Once we have guaranteed same velocities in the potential region,
we have to check the probability of transmission (or equivalently the
probability of reflection). A simple calculation (see section III) shows
that
\[
\begin{array}{lclcl}
P_{c,\r} & \approx & \displaystyle{\left(
\frac{\sqrt{E_{\0}}-\sqrt{E_{\0}-W_{\0}}}{
\sqrt{E_{\0}}-\sqrt{E_{\0}-W_{\0}}}\,\right)^{\2}} & \approx &
2.01\,\times\,10^{-\2}\,\,,\\
P_{q,\r} & \approx & \displaystyle{\frac{\left[\sqrt{E_{\0}}-
\left(E_{\0}^{^{\2}}-V_{\0}^{^{\2}}\right)^{\1/\4}\right]^{\2}}{
E_{\0}-\sqrt{E_{\0}^{^{\2}}-V_{\0}^{^{\2}}}}} & \approx &
2.58\,\times\,10^{-\3}\,\,.
\end{array}
\]
This gives an interesting answer to the old question concerning the
possibility to fit a quaternionic potential by a complex one and
consequently to never recognize deviations from standard quantum mechanics.
The study presented for a very simple potential, i.e. the potential step,
shows that incoming particles with a given energy spectrum behave
differently if they are diffused by a complex or a pure quaternionic
potential step. This result, obtained by numerical calculations and
interpreted by our gaussian analytical approximations, cannot be seen in
the plane wave analysis. In such a limit, we only have reflection and
transmission probabilities, and the quaternionic case can {\em always} be
interpreted in terms of a complex potential step which appropriately fits
the quaternionic probabilities. This paper stimulates further
investigations, for example  it should be interesting to study, within the
wave packet formalism, diffusion and tunnelling by complex and pure
quaternionic potential barrier which, surely, represents a more realistic
potential  to be tested by an experimental analysis. In the case of the
potential barrier, it is of great interest to examine the tunnelling
phenomena, in particular, in a forthcoming paper we aim to analyze in
detail the Hartman effect\cite{HAR,RECAMI} in the presence of a pure
quaternioic potential. The possibility to treat the potential barrier
problem as a two-step problem\cite{MPLA19,EPJC46} gives a more solid
interest to this paper, what it could appear an academic quantum mechanical
exercise play a fundamental role in the understanding the behavior of wave
packets in the presence of a more complicate potential. The analysis done
in this paper explicitly shows that a {\em qualitative} difference exists
between complex and quaternionic quantum mechanics.

\newpage

\begin{figure}[hbp]
\vspace*{-1cm} \hspace*{-2.5cm}
\includegraphics[width=19cm, height=22cm, angle=0]{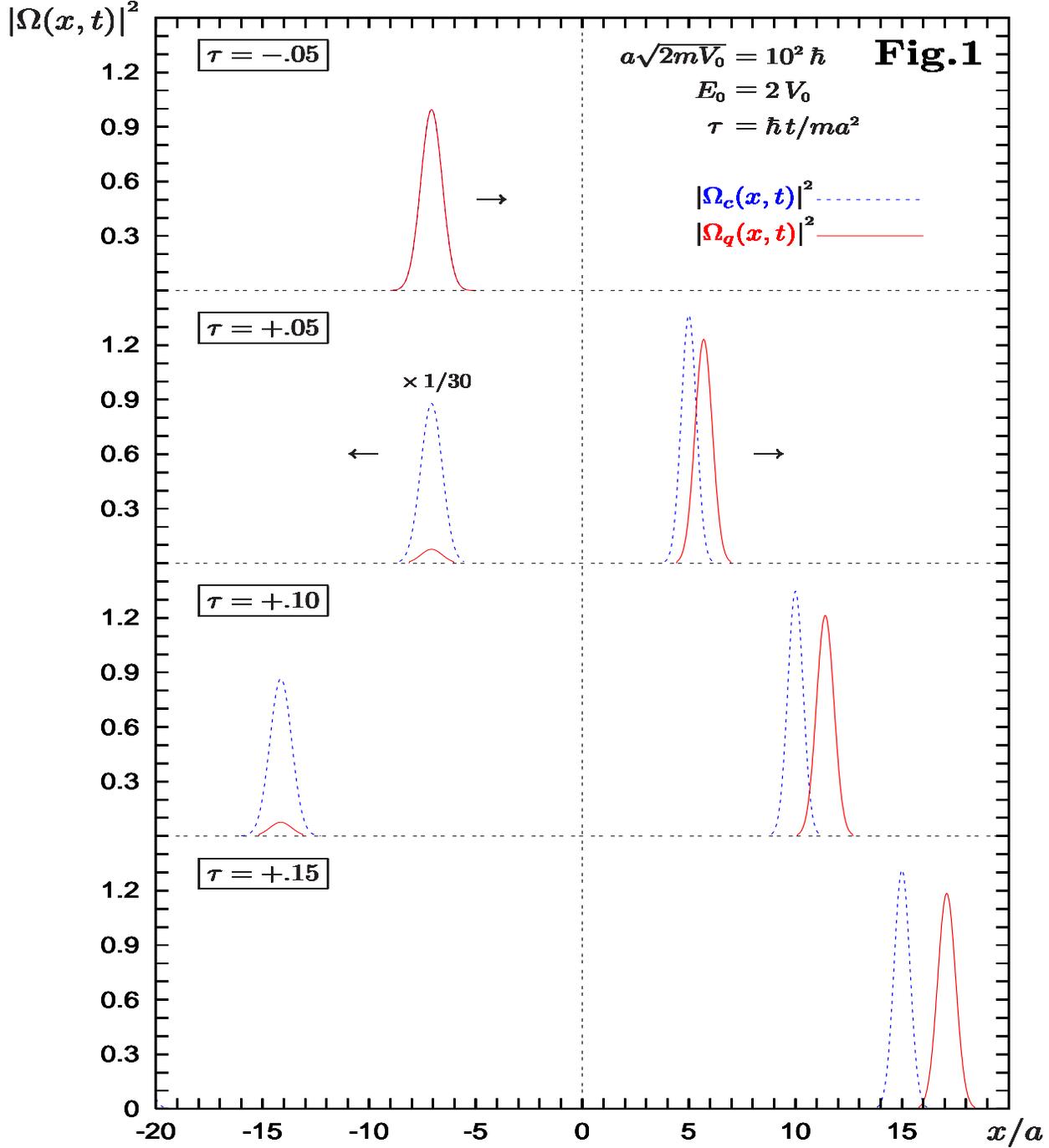}
\vspace*{-1cm} \caption{Diffusion ($E_{\0}=2\,V_{\0}$) of a wave packet at
a complex and pure quaternionic potential step. The complex
($V_{\0}=V_{\1}$) and quaternionic
($V_{\0}=\sqrt{V_{\2}^{^{\2}}+V_{\2}^{^{\2}}}$) potentials act for positive
$x$. For negative time ($\tau<0$) only the incident packet is present and
it moves towards the step. After a certain time, we find four packets. The
reflected packets, $|\Omega_{c,\i}(x,t)|^{\2}$ and
$|\Omega_{q,\i}(x,t)|^{\2}$, are returning to the left moving with the same
velocity. The transmitted packets, $|\Omega_{c,\t}(x,t)|^{\2}$ and
$|\Omega_{q,\t}(x,t)|^{\2}$, propagate towards  the right and move with
different velocities.}
\end{figure}

\end{document}